# Lightweight Speaker Verification Using Transformation Module with Feature Partition and Fusion

Yanxiong Li, Zhongjie Jiang, Qisheng Huang, Wenchang Cao, and Jialong Li

*Abstract*—Although many efforts have been made on decreasing the model complexity for speaker verification, it is still challenging to deploy speaker verification systems with satisfactory result on low-resource terminals. We design a transformation module that performs feature partition and fusion to implement lightweight speaker verification. The transformation module consists of multiple simple but effective operations, such as convolution, pooling, mean, concatenation, normalization, and element-wise summation. It works in a plug-and-play way, and can be easily implanted into a wide variety of models to reduce the model complexity while maintaining the model error. First, the input feature is split into several low-dimensional feature subsets for decreasing the model complexity. Then, each feature subset is updated by fusing it with the inter-feature-subsets correlational information to enhance its representational capability. Finally, the updated feature subsets are independently fed into the block (one or several layers) of the model for further processing. The features that are output from current block of the model are processed according to the steps above before they are fed into the next block of the model. Experimental data are selected from two public speech corpora (namely VoxCeleb1 and VoxCeleb2). Results show that implanting the transformation module into three models (namely AMCRN, ResNet34, and ECAPA-TDNN) for speaker verification slightly increases the model error and significantly decreases the model complexity. Our proposed method outperforms baseline methods on the whole in memory requirement and computational complexity with lower equal error rate. It also generalizes well across truncated segments with various lengths.

## I. Introduction

SPEAKER recognition is the task for recognizing a person based on his/her voices, which is one major development direction of biometrics [1]. With the wide application of voice-enabled terminals (e.g., smart watches, smart phones, smart earphones), speaker recognition has become a crucial technology in many practical scenarios of voice biometrics, such as criminal investigation [1], financial security [2]. For instance, law enforcement agencies usually need to confirm whether the recorded voices are spoken by the claimed person or not, which is a key cue for solving criminal cases. In past decades, judges, detectives, lawyers, and courts intended to utilize speaker recognition as a powerful tool to identify criminals [1]. In the application of financial security, speaker recognition is one of the main ways for remote identity authentication, which is utilized to protect the security of the funds in the accounts [2]. In addition, speaker recognition is also a basic or an indispensable part for implementing other speech related tasks, such as speaker clustering [3], speaker diarization [3], multi-speaker speech recognition [4], and speaker tracking [5].

The task of speaker recognition generally includes two sub-tasks: speaker identification and Speaker Verification (SV) [6]. The task of speaker identification is to decide which enrolled person utters a given voice from a set of known persons, while the task of SV is to reject or accept the identity claim of a person based on his/her utterance. The work in this paper concentrates on the task of SV only. Concretely, we propose a method for lightweight SV using a Transformation Module (TM) with feature partition and fusion.

The rest of this paper is organized as follows. Sections II and III present related works and our contributions, respectively. Section IV describes the proposed method in detail. Section V gives the experiments and discussions, and the conclusions are drawn in Section VI.

## II. Related Works

Many studies have been conducted on the task of SV [7]-[13]. The goal of these previous works was to tackle two main technical problems. The first problem is to effectively learn a feature with powerful representational capability. The second problem is to construct a classifier with strong classification ability.

Many hand-crafted features (shallow-model based features) were designed to represent the time-frequency characteristics of different speakers, including Mel-frequency cepstral coefficients [14], constant Q cepstral coefficients [15], linear prediction coding coefficients [14], eigenvoice-motivated vectors [16], and I-vector [17], [18]. They were proposed for specific conditions and thus had poor generalization ability for other situations. Furthermore, they could not effectively characterize the differences of deep-level properties among different speakers, since they were obtained based on shallow models rather than deep ones. Afterwards, the deep-model based features were learned using different kinds of deep neural networks. These features mainly included the X-vector learned by a Time-Delay Neural Network (TDNN) [19]-[23] or an Emphasized Channel Attention, Propagation and Aggregation in TDNN (ECAPA-TDNN) [24]; the R-vector learned by a Residual Network with 34 layers (ResNet34) [25]; the S-vector

This work was supported by national natural science foundation of China (62371195, 62111530145, 61771200), Guangdong basic and applied basic research foundation (2021A1515011454), international scientific research collaboration project of Guangdong (2021A0505030003, 2023A0505050116), and the Guangdong Provincial Key Laboratory of Human Digital Twin (2022B1212010004).

All authors of this paper are with School of Electronic and Information Engineering, South China University of Technology, Guangzhou, China. The corresponding author is Dr. Yanxiong Li (eeyxli@scut.edu.cn).

learned by a Transformer [26]. In addition, other kinds of neural networks were adopted to learn deep embeddings [27]-[35], such as temporal dynamic convolutional neural network [31], Attentive Multi-scale Convolutional Recurrent Network (AMCRN) [33], Siamese neural network [34], and long short-term memory network [35].

Besides the works on feature learning (extraction), many efforts were made on the construction of back-end classifiers for SV. The typical classifiers mainly included the Cosine Distance (CD) [1], Probabilistic Linear Discriminant Analysis (PLDA) [36], [37], and deep neural network [38].

Although the SV methods proposed in these aforementioned works obtained low Equal Error Rate (EER), the reduction of model complexity (including both computational complexity and memory requirement) was not explicitly considered. Accordingly, the model complexity of these methods was very high, and thus they cannot be directly deployed on the terminals with limited resources. To reduce the model complexity for implementing lightweight SV, some recent works were done on the design of lightweight models or the compression of model parameters without significantly increasing the model error. These efforts include the design of computationally-efficient convolution, the manual or automatic design of better model architectures, and the construction of small student model, which are briefly summarized as follows.

Inspired by the success of the Depth-wise Separable Convolutions (DSC) adopted in the MobileNet for mobile vision applications [39], both the SpeakerNet [40] and the AM-MobileNet1D [41] were designed for realizing lightweight SV. The SpeakerNet mainly consists of residual blocks with one-dimensional (1D) DSC and has 5 million parameters, while the AM-MobileNet1D occupies 11.6 megabytes of memory. Although the DSC operations significantly reduced the model complexity, they caused the increase of the error. What is more, the tradeoff between the reduction of the model complexity and the increase of the model error was generally hard to achieve in practice.

The model architectures were manually designed to reduce the model complexity for implementing lightweight SV [42]-[44]. In the work of [42], the original trunk of the SincNet [45] was replaced by a lightweight trunk with 2.8 million parameters for reducing the model complexity. Lee et al. [44] designed a hyperbolic ResNet for lightweight application. Their model learned more compact deep embeddings with equivalent error. In short, the size of the models whose architectures were manually designed generally reached the level of millions of parameters.

Additionally, some researchers applied the techniques of Knowledge Distillation (KD) [46], [47] and Neural Architecture Search (NAS) [48] to implement lightweight SV [49]-[52]. In the work of [49], the strategy of teacher-student training was proposed for text-independent SV, and competitive error rate with 88-93% smaller models was obtained. Lin et al. [50] designed a framework with asymmetric structure, in which a large model was used for enrollment and a small model was used for verification. The generated small model achieved competitive results with 11.6 million floating-point operations per second. Recently, the NAS technique was applied to design an efficient model (termed EfficientTDNN) which obtained satisfactory error rate with low computational complexity [51].

In short, these previous methods produced a small model for verification. However, the KD based methods had to construct a large model for training the small model, and the NAS based methods required to search the appropriate model with a lot of different architecture settings. As a result, the training expense and whole procedure for implementing lightweight SV are actually not light.

III. OUR CONTRIBUTIONS

Based on the introductions in Section II, it can be concluded that a lot of efforts have been made on solving the problems of both general SV and lightweight SV. However, there are some shortcomings in prior works. First, there is still room to reduce the model complexity. For example, the number of model parameters in prior methods is almost over one million for obtaining lower EER scores. Second, the structure of the original model needs to be significantly modified or a new model has to be designed for implementing lightweight SV. For example, it is required to redesign the model structure, replace some components, search for many model architectures, or pre-train a large model. These operations are not simple, and various operations are required to lighten models with different architectures. That is, each one of the previous methods can be only applicable to one specific model in practice, and thus lacks generality.

To overcome the aforementioned shortcomings in the prior works, we design a TM to execute the operations of both feature partition and feature fusion for realizing lightweight SV. What is more, we propose a method for lightweight SV by implanting the TM into models with different architectures. Experimental results on three different evaluation sets indicate that implanting the proposed TM into three state-of-the-art models for SV obtains equivalent EER scores or slightly-lower EER scores with remarkable reduction of the model complexity. Compared with the state-of-the-art methods of lightweight SV, our proposed method can obtain lower EER scores and model complexity. What is more, the proposed method also obtains satisfactory results when evaluated on truncated segments with different durations. In a word, main contributions of this work are briefly summarized as follows.

1. To reduce the model complexity and obtain the transformed features with powerful representational capability, we design a TM to perform feature partition and feature fusion. The proposed TM is composed of some simple but effective operations, such as convolution, pooling, concatenation, mean, normalization, and element-wise summation. It can be easily implanted into the original models with various architectures for reducing the model complexity without replacing any components or changing the architectures of the original models. That is, it can work in a plug-and-play way and can be utilized to reduce the complexity of many types of models for realizing lightweight SV. To the best of our knowledge, the proposed TM is novel and is not adopted in prior works.
2. We propose a method for lightweight SV by placing the TM in front of different blocks of the original model. The proposed method is a general solution for lightening existing models for realizing lightweight SV, whereas each one of all previous methods is a specific solution for lightweight SV and lacks generality. In addition, we comprehensively evaluate the effectiveness of the proposed method, and compare it with the state-of-the-art methods on three different evaluation sets under various experimental conditions. Experimental results indicate that our proposed method basically has advantages over the baseline methods in both model error and model complexity under the same experimental conditions.

## IV. METHOD

Fig. 1 illustrates the implantation of the TM into the model for SV, where the TM is placed in front of the block of the model. The TM consists of two blocks: feature partition and feature fusion. First, the input feature $F$ ($T$ frames and $N$ dimensions per frame) is segmented into $J$ feature subsets $F_i$ ($T$ frames and $L$ dimensions per frame, $1 \leq i \leq J$) by the feature partition block. Afterwards, each $F_i$ is updated to $F_i'$ by the feature fusion block. Finally, each $F_i'$ is fed into the block of the model to produce the transformed feature subset $F_i''$.

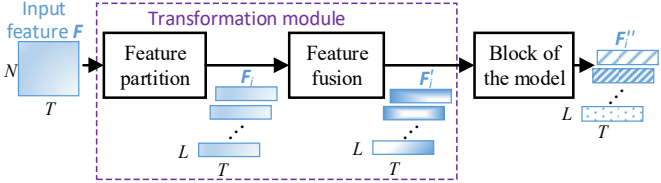

Fig. 1 The schematic diagram of implanting the transformation module into the model. The feature $F$ is converted into feature subsets $F_i$, $F_i'$ and $F_i''$ in turn.

The motivation for implanting the TM into the model for SV is based on two considerations. First, we do not want to change the architecture of the original SV model. The implantation of the TM can be implemented by placing the TM in front of each block (layer) of the model. Second, implanting the TM into the model not only reduces the model complexity (by feature partition), but basically does not increase the model error (by feature fusion). The feature partition block splits the input feature $F$ into $J$ feature subsets $F_i$ that will be independently fed into the block of the model after feature fusion. That is, after the implantation of the TM, the input of each block of the model becomes low-dimensional feature subsets rather than the high-dimensional input feature. Therefore, the number of parameters of each block of the model can be reduced when each feature subset is fed into the model. However, the correlational information (namely cross sub-band dependency or global contextual spectral information) between feature subsets $F_i$ cannot be utilized if each $F_i$ is independently fed into the model. That is, the model cannot see the complete feature and thus cannot capture global contextual spectral information when each $F_i$ is independently fed into the model. As a result, the representational capability of the transformed feature will be weakened. To make the model see the complete feature and capture the correlational information between $J$ feature subsets $F_i$, we design a block of feature fusion that enables feature subsets $F_i$ to interact with each other. After feeding $F_i$ into the feature fusion block, we obtain feature subsets $F_i'$. Each $F_i'$ contains the correlational information between different $F_i$, rather than a fragmented feature subset. The correlational information is beneficial for enhancing the representational capability of the learned feature. Therefore, the model error for SV is expected to be maintained when the feature subsets $F_i'$ are adopted as the input of the model.

### A. Transformation Module

Fig. 2 shows the framework of the TM. The feature partition block is composed of one operation of feature division, while the feature fusion block comprises the operations of 1D convolution, mean pooling, mean calculation, concatenation, Z-score normalization, and element-wise summation.

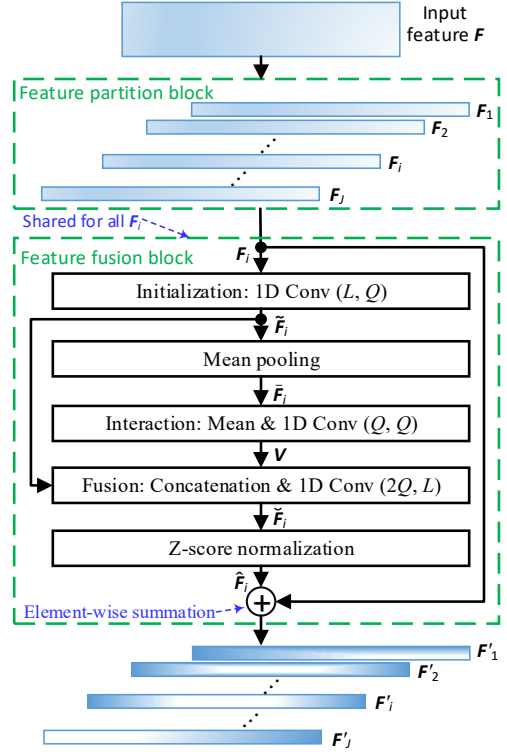

Fig. 2 The framework of the proposed transformation module for feature partition and fusion. 1D Conv ($L$, $Q$) denotes one-dimensional convolution with $L$ input channels and $Q$ output channels.

As shown in Fig. 2, the input feature $F \in \mathbb{R}^{N \times T}$ is first fed into the block of feature partition, and is split into $J$ feature subsets $F_i \in \mathbb{R}^{L \times T}$ along the channel-dimension of deep transformed feature or along the frequency-dimension of audio feature. $T$ denotes total number of frames. $N$ and $L$ stand for the dimension of the feature and feature subsets, respectively. When adjacent feature subsets overlap to varying degrees, the value of $J$ will change accordingly. For example, $N$=80, $L$=20, if all adjacent feature subsets do not overlap, then $J$=4. If there is 50% overlap between all adjacent feature subsets, then $J$=7. After feature partition, each feature subset is fragmented, and thus does not contain the correlational information between $J$ feature subsets. The correlational information is originally contained in the complete feature.

To capture the correlational information between $J$ feature subsets $F_i$, we design a block of feature fusion. Each feature subset $F_i$ is first fed into an initialization layer for producing the feature subset $\widetilde{F}_i \in \mathbb{R}^{Q \times T}$. The initialization layer is to conduct a one-dimensional convolution whose kernel size is 1, with $L$ input channels and $Q$ output channels. Afterwards, each feature subset $\widetilde{F}_i$ is transformed into $\overline{F}_i \in \mathbb{R}^{Q \times T}$ by one mean pooling layer. The feature subset $\overline{F}_i$ is fed into an interaction layer for

generating the interacted vector $V \in \mathbb{R}^{Q \times T}$ which is defined by

$$V = \mathcal{R}\left(\frac{1}{J}\sum_{i=1}^{J} \overline{F}_i\right), \quad (1)$$

where $\mathcal{R}(\cdot)$ denotes a one-dimensional convolution whose kernel size is 1, with $Q$ input channels and $Q$ output channels. The interacted vector $V$ is transformed from the mean vector of $J$ feature subsets and the output of one-dimensional convolution, so it contains the correlational information between $J$ feature subsets. Afterwards, each feature subset $\widetilde{F}_i$ and the interacted vector $V$ are fed into a fusion layer which is composed of the operations of concatenation and one-dimensional convolution. In the fusion layer, the feature subset $\widetilde{F}_i$ and the interacted vector $V$ are concatenated and then transformed by the one-dimensional convolution for producing the feature subset $\breve{F}_i \in \mathbb{R}^{L \times T}$ which is obtained by

$$\breve{F}_i = \mathcal{H}([\widetilde{F}_i, V]), \quad (2)$$

where $\mathcal{H}(\cdot)$ represents a one-dimensional convolution whose kernel size is 1, with $2Q$ input channels and $L$ output channels; and $[\cdot, \cdot]$ denotes an operation of concatenation. We obtain the feature subset $\widehat{F}_i = \{\hat{f}_i^{l,t}\} \in \mathbb{R}^{L \times T}$ after feeding the feature subset $\breve{F}_i = \{\breve{f}_i^{l,t}\}$ into a Z-score normalization layer, where $1 \le l \le L$ and $1 \le t \le T$. Namely, the $\hat{f}_i^{l,t}$ is obtained by

$$\hat{f}_i^{l,t} = \frac{\breve{f}_i^{l,t} - \mu_i^t}{\sigma_i^t}, \quad (3)$$

where $\hat{f}_i^{l,t}$ and $\breve{f}_i^{l,t}$ represent the $l$-th element of the $t$-th frame in the feature subsets $\widehat{F}_i$ and $\breve{F}_i$, respectively; $\mu_i^t$ and $\sigma_i^t$ denote mean and standard deviation of the $t$-th frame in the feature subset $\breve{F}_i$, respectively. Finally, the feature subsets $\widehat{F}_i$ and $F_i$ are element-wisely summed to generate the feature subset $F_i' \in \mathbb{R}^{L \times T}$.

In summary, the function of the proposed TM is to transform the input feature $F \in \mathbb{R}^{N \times T}$ into $J$ feature subset $F_i' \in \mathbb{R}^{L \times T}$. Moreover, each feature subset $F_i'$ contains the correlational information between $J$ feature subsets $F_i$. The transformation process above is defined by

$$F_i' = \mathcal{F}(F), \quad (4)$$

where $\mathcal{F}(\cdot)$ is the conversion function of the proposed TM.

*B. Implanting the Transformation Module into Models*

To demonstrate the effectiveness of the proposed TM for reducing the model complexity, we implant the TM into three state-of-the-art models for SV, namely ECAPA-TDNN [24], ResNet34 [25], and AMCRN [33]. Specifically, we place the proposed TM in front of each block in the frame-level module rather than the utterance-level module of the model. Main reasons for doing so are as follows.

First, the computational load of the blocks in the frame-level module is much heavier than that of the blocks in the utterance-level module, because the frame-level module is the main part of the model. Accordingly, implanting the proposed TM into the blocks of the frame-level module can significantly reduce the complexity of the model.

Second, the semantic abstraction degrees of the features learned by the blocks in the frame-level module are much lower than that obtained in the utterance-level module, because the utterance-level module is located in the deeper position of the model. The outputs of both feature partition and feature fusion from the frame-level module are expected to cause less damage to the abstract semantic information which has influence on the representational capability of the learned features.

The implantation of the TM into the AMCRN is shown in Fig. 3. The AMCRN model is composed of the blocks in black solid-line boxes, whose detailed information is introduced in [33]. The proposed TM (as illustrated in green solid-line box) is placed in front of each block of the frame-level module in the AMCRN model.

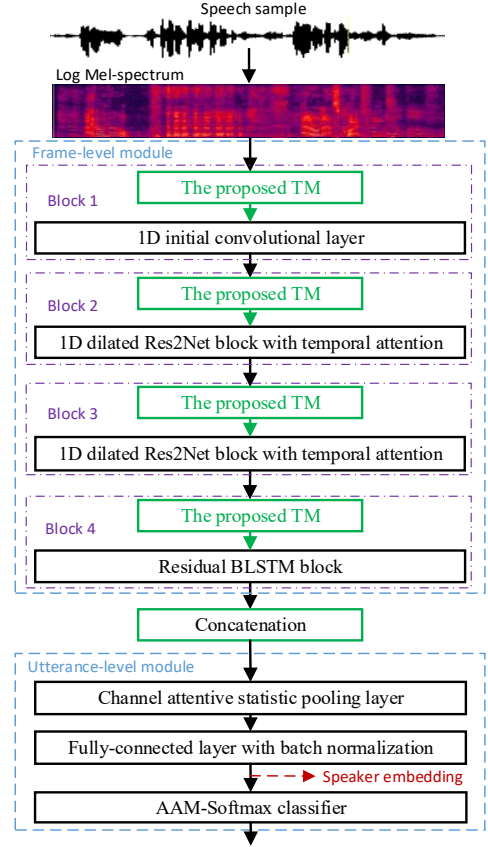

Fig. 3 The implantation of the TM into the AMCRN. BLSTM: bidirectional long short-term memory; AAM-Softmax: additive angular margin Softmax.

The AMCRN model after the implantation of the TM works as follows. First, audio feature of Log Mel-spectrum [53] is extracted from each speech sample and fed into the TM for splitting the Log Mel-spectrum feature into feature subsets and fusing these feature subsets. According to the extraction process of Log Mel-spectrum in [53], Log Mel-spectrum is actually the feature of FBank (Filter-Bank). To facilitate the visualization of feature partition, this audio feature is called Log Mel-spectrum here. The feature subsets that are output from the proposed TM are independently and sequentially input into the 1D initial convolutional layer (in Block 1) for further processing. Next, the transformed feature subsets are fed into the proposed TM (in Block 2) and then transformed by the 1D dilated Res2Net block with temporal attention. Similarly, the transformed feature subsets are sequentially processed by the operations in Block 3 and Block 4. Afterwards, these feature subsets are spliced together by the Concatenation layer (green solid-line box) to form a complete feature which is input into the utterance-level module for further transformation. Finally, the speaker embedding is produced from the Fully-connected layer with batch normalization for scoring by the CD or PLDA.

The implantations of the proposed TM into the ResNet34 and ECAPA-TDNN models are illustrated in Fig. 4 (a) and Fig. 4 (b), respectively. The ResNet34 and ECAPA-TDNN models are composed of layers (or blocks) in black solid-line boxes in Fig. 4 (a) and Fig. 4 (b), respectively. The workflows of these two models after the implantation of the proposed TM are similar to that of the AMCRN model. The ECAPA-TDNN and ResNet34 models are described in detail in [24] and [25], respectively. To effectively apply the proposed TM for the lightweighting of ResNet34 model, all convolutions in the ResNet34 model for the proposed method of ResNet34-TM are set to 1D convolutions in the experiments.

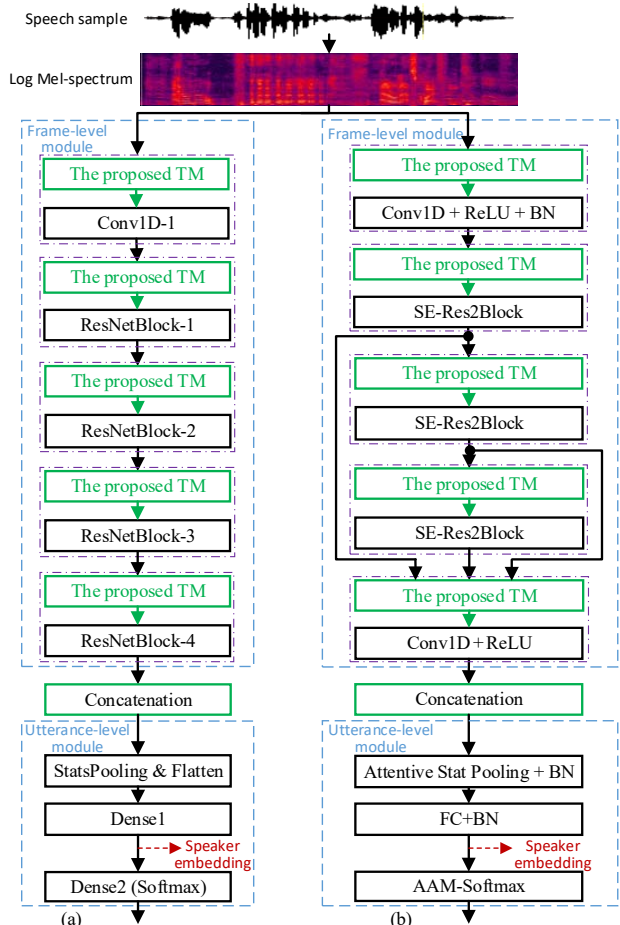

Fig. 4 The implantations of the TM into: (a) the ResNet34 and (b) the ECAPA-TDNN. Conv1D: one-dimensional convolution; ReLU: Rectified Linear Unit; BN: batch normalization; StatsPooling: statistics pooling; SE-Res2Block: Squeeze-and-Excitation Res2Net block; FC: fully-connected.

## V. EXPERIMENTS AND DISCUSSIONS

In this section, experimental datasets and setups are first introduced. Then, we perform ablation experiments for the proposed method. Next, we compare our proposed method to the baseline methods in the error and complexity. Finally, we analyze the robustness of three representatives of the proposed method on truncated testing segments with various lengths.

### A. Experimental Datasets

Experimental datasets were selected from two large-scale speech corpora: the VoxCeleb1 [54] and the VoxCeleb2 [55]. These two speech corpora are publicly available for research and widely utilized for speaker verification in prior works.

The VoxCeleb1 consists of 153,516 speech samples spoken by 1,251 speakers. Its duration is 352 hours in total. Speech samples of the VoxCeleb1 are collected from the sound tracks of the videos on YouTube. The gender ratio of the VoxCeleb1 is basically balanced, with 45% of female speakers. Speakers contained in the VoxCeleb1 span a wide range of different ages, professions, ethnicities, and accents. The contents of the speech samples are mainly composed of speeches, interviews, excerpts from professionally shot multimedia. There are several kinds of real-world noise in the speech samples, such as background voices (e.g., chatter, laughter), room reverbs, sampling equipment noise.

The VoxCeleb2 is an extended speech corpus of the VoxCeleb1, and consists of 1,128,246 speech samples spoken by 6,112 speakers. Its duration is 2,442 hours in total. The gender ratio of the VoxCeleb2 is roughly balanced with 61% of male speakers. The speakers cover a great range of professions, accents, ethnicities, and ages. Its speech samples are degraded with background noise, such as overlapping speech, chatter, laughter, and varying room acoustics.

Experimental data was chosen from the VoxCeleb1 and the VoxCeleb2, and was composed of three datasets, namely VoxCeleb2-dev, VoxCeleb1-dev, and VoxCeleb1-test. These three datasets were split by the data organizer of the VoxCeleb [56] and widely adopted in prior works [54], [55]. The training data was chosen from the VoxCeleb2-dev, and contained 5,994 speakers and 1,092,009 speech samples in total. The testing data came from the VoxCeleb1-dev and the VoxCeleb1-test, and had three different evaluation sets: Vox1-O, Vox1-E and Vox1-H, where O, E, and H denote Original, Extended, and Hard data subsets of the VoxCeleb1, respectively. These three different evaluation sets were split by the data organizer of the VoxCeleb [56]. The proportions of training and validation subsets that were contained in the training data were 95% and 5%, respectively. The average duration of each speech sample is basically 8 seconds. Table I gives the detailed information of the data adopted in the experiments of this paper.

TABLE I
THE DETAILED INFORMATION OF EXPERIMENTAL DATA

|  | Names | Dataset sources | #Speakers | Data size |
|---|---|---|---|---|
| Training | / | VoxCeleb2-dev | 5,994 | 1,092,009 samples |
| Testing | Vox1-O | VoxCeleb1-test | 40 | 37,611 trails |
|  | Vox1-E | VoxCeleb1-dev | 1,211 | 579,818 trails |
|  | Vox1-H | VoxCeleb1-dev | 1,211 | 550,894 trails |

### B. Experimental Setup

The experiments are performed on a computing machine whose main configurations include: an Intel CPU i7-6700 with 3.10 GHz, a RAM of 64 GB, and a NVIDIA 1080TI GPU. All methods are implemented on the deep learning toolkit of the PyTorch. Two different indices are adopted to assess the performance of the methods (models), namely error and complexity. The metric of EER [57] is utilized to measure the error of different methods, which has been commonly used in prior works. The EER is defined as the rate at which acceptance error is equal to rejection error. The lower the score of the EER is, the lower the error of the method is. The complexity index contains two metrics, namely Parameter Number (PN) and Multiply-ACcumulate operations (MACs). The PN and MACs are adopted to measure the memory requirement and the computational complexity of different methods, respectively. The MACs is defined as the total amount of multiplication and addition operations of the model. The PN is defined as the total amount of parameters of the model. The lower the value of MACs is, the lower computational complexity of the method is. Similarly, the lower the value of PN is, the lower memory requirement of the method is. The MACs and PN have been widely used to measure the model complexity in prior works, and thus are also adopted as the evaluation metrics here.

The audio feature adopted as the input of the model is the 80-dimensional Log Mel-spectrum. The Log Mel-spectrum is extracted from each speech sample with frame length of 25 ms and frame shift of 10 ms. It is subject to short-time cepstral mean variance normalization using a 3-second sliding window. To expand the diversity of the training data, we set a probability of 0.6 for data augmentation. We randomly add the background noise, reverberation and speed change on the original speech samples when they are chosen to be augmented. All the background noise and reverberation audios are selected from two public noise datasets MUSAN [58] and RIR [59]. The speed change rate is set to a value ranging from 95% to 105% of the original audios. Both the original speech samples and the augmented speech samples are adopted to train the models.

The structure settings of the three models (AMCRN, ResNet34, and ECAPA-TDNN) are the same as those in the corresponding references. All models are trained for 50 epochs using the cosine annealing learning rate [60] with 10 warm-up steps from $10^{-3}$ to $10^{-8}$ in conjunction with the Adam optimizer [61]. We use the AAM-Softmax [62] with a margin of 0.2 and Softmax pre-scaling of 30 to train all models. Early stopping technique is adopted when the loss of model validation no longer decreases for 10 epochs. The mini-batch size for training is equal to 512. After each epoch on the training subset, the trained model is assessed on the validation subset. The model with the lowest validation loss is used as the final model. After completing the model training, enrollment data and testing data are input into the model to produce the speaker embedding for each speech sample. The dimension of speaker embedding is set to 256. Finally, the CD is used as the back-end classifier for scoring on each testing trail.

The dimension of Log Mel-spectrum is 80 in the experiments, and thus the value of $N$ is equal to 80. In addition, $Q$, the number of channels of 1-dimensional convolution, is set to $2L$. The settings of $J$ will be discussed in ablation experiments, which has direct impact on the model complexity.

*C. Ablation Experiments*

In this section, we conduct three ablation experiments. First, we discuss the influence of the settings of main parameters (i.e., $J$, $L$, $N$) of the proposed TM on the error and complexity of the models. Second, we analyze the impact of the overlap between different feature subsets on the model complexity and error. Third, we qualitatively analyze the impact of feature fusion on the representational capability of feature subsets. Without loss of generality, we adopt the AMCRN as the model and use the Vox1-O as the testing data in the experiments of this section.

**Main parameter settings of the TM**

In this ablation experiment, the values of $J$ range from 1 to 16 in steps of the power of 2, namely 1, 2, 4, 8, and 16. The values of $L$ range from 80 to 5, namely 80, 40, 20, 10, and 5. There is no overlap between $J$ feature subsets in this experiment. In the AMCRN model, the number of input channels of the 1D dilated Res2Net block (denoted as $C$) and the number of neurons of the residual BLSTM block (denoted as $B$) are two main factors that determine the model complexity. To obtain both lower error and lower complexity, the values of both $C$ and $B$ need to be appropriately configured according to the dimensions of feature subsets. When $L$ takes values from 80 to 5, we give three to four settings of $B$ and $C$ for each value of $L$. The values of $B$ and $C$ range from 512 to 16. Table II lists the scores of EER, PN, and MACs that are obtained by the AMCRN model with various values of $B$ and $C$ on the Vox1-O.

TABLE II
IMPACTS OF MAIN PARAMETER SETTINGS OF THE TM ON THE ERROR AND COMPLEXITY OF THE AMCRN MODEL WHEN EVALUATED ON THE VOX1-O

| $J$ | $L$ | $N$ | $C$ | $B$ | EER (%) | PN | MACs |
|---|---|---|---|---|---|---|---|
| 1 | 80 | 80 | 512 | 512 | 1.464 | 11.38 M (100%) | 0.840 G (100%) |
|   |    |    | 256 | 256 | 1.898 | 3.59 M (31.5%) | 0.250 G (29.8%) |
|   |    |    | 128 | 128 | 3.734 | 970.01 K (8.5%) | 0.081 G (9.6%) |
| 2 | 40 | 80 | 256 | 256 | 1.470 | 6.45 M (56.7%) | 0.276 G (32.9 %) |
|   |    |    | 128 | 128 | 1.896 | 1.76 M (15.5%) | 0.093 G (11.1%) |
|   |    |    | 64  | 64  | 2.892 | 571.67 K (5.0%) | 0.036 G (4.3%) |
| 4 | 20 | 80 | 256 | 256 | **1.459** | 6.93 M (60.9%) | 0.347 G (41.3%) |
|   |    |    | 128 | 128 | 1.728 | 2.01 M (17.7%) | 0.128 G (15.2%) |
|   |    |    | 64  | 64  | 2.624 | 644.89 K (5.7%) | 0.053 G (6.3%) |
| 8 | 10 | 80 | 256 | 256 | 1.461 | 7.98 M (70.1%) | 0.501 G (59.6%) |
|   |    |    | 128 | 128 | 1.655 | 2.53 M (22.2%) | 0.205 G (24.4%) |
|   |    |    | 64  | 64  | 2.325 | 902.05 K (7.9%) | 0.091 G (10.8%) |
|   |    |    | 32  | 32  | 2.981 | 360.79 K (3.2%) | 0.043 G (5.1%) |
| 16 | 5 | 80 | 128 | 128 | 1.592 | 3.58 M (31.5%) | 0.362 G (43.1%) |
|    |   |    | 64  | 64  | 2.044 | 1.43 M (12.6%) | 0.170 G (20.2%) |
|    |   |    | 32  | 32  | 2.757 | 622.63 K (5.5%) | 0.082 G (9.8%) |
|    |   |    | 16  | 16  | 3.205 | **289.02 K (2.5%)** | **0.040 G (4.8%)** |

$J$: number of feature subsets; $L$: dimension of feature subsets; $N$: dimension of input feature; $C$: number of input channels of the 1D dilated Res2Net block of the AMCRN; $B$: number of neurons of the residual BLSTM block of the AMCRN. K: Kilo; M: Million; G: Giga.

Based on the results presented in Table II, the following three observations can be obtained.

First, when the value of $J$ equals to 1 (i.e., $L=N$), the input feature is not split into feature subsets and thus the proposed TM is not used for reducing the model complexity. In this condition, when the values of both $B$ and $C$ are set to 512, the EER score of 1.464%, the PN value of 11.38 M, and the MACs value of 0.840 G are obtained. The values of 11.38 M and 0.840 G are used as the benchmark (represented by 100% in Table II) for reducing the complexity of the AMCRN model. Although the complexity of the model can be reduced by decreasing the values of $B$ and $C$ without using the proposed TM, the model error increases greatly. For example, when the values of $B$ and $C$ decrease from 512 to 128, the values of PN and MACs decrease from 11.38 M to 970.01 K (8.5% of 11.38 M), and from 0.840 G to 0.081 G (9.6% of 0.840 G), respectively. However, the scores of EER increase from 1.464% to 3.734%.

Second, with the increase the value of $J$, the model complexity is remarkably reduced while keeping the model error basically unchanged or even better. That is, when the EER score is basically equal to or even lower than 1.464%, the values of both PN and MACs are significantly reduced compared with 11.38 M and 0.840 G, respectively. For example, when $J$ increases to 4, and the values of both $C$ and $B$ are set to 256, the lowest EER score of 1.459%, the PN value of 6.93 M

(60.9% of 11.38 M), and the MACs value of 0.347 G (41.3% of 0.840 G) are produced. This result indicates that not only the model complexity is significantly reduced, but also the model error decreases after implanting the TM into the model. In addition, when the model complexity is close, the model error after the implantation of the TM is lower than that of the original model (without the implantation of the TM). For example, when $J=8$, $B=C=64$, the model complexity (i.e., PN = 902.05 K and MACs = 0.091 G) is close to that (i.e., PN = 970.01 K and MACs = 0.08 G) of the original model in which $J=1$, $B=C=128$. However, the EER score of 2.325% is obtained by the model after the implantation of the TM, which is lower than that (3.734%) produced by the original model.

Third, when $J$ increases to a certain value (e.g., 16), the model complexity decreases significantly, but the model error gradually increases and is worse than that of the original model. For example, when $J=16$ and $B=C=16$, the PN value decreases to 289.02 K (only 2.5% of 11.38 M) and the MACs value decreases to 0.040 G (only 4.8% of 0.840 G), but the EER score increases to 3.205% (higher than 1.464%). The possible reason is described as follows. With the increase of $J$, the dimension of feature subsets will become smaller and smaller. When the dimension of feature subsets is reduced to a certain value (e.g., $L=5$), the correlational information between these $J$ feature subsets is so damaged that the fusion block of the TM cannot recover it. As a result, the error obtained by the model after the implantation of the TM cannot be smaller than that obtained by the original model when $J$ exceeds a threshold (e.g., 16).

In summary, this experiment investigates the impact of the TM implantation on the model error and model complexity by setting different values of $J$, $N$, $C$ and $B$. The experimental results indicate that the model complexity is greatly reduced when the model error is basically unchanged or even becomes lower after the implantation of the TM.

**Impact of the overlap between feature subsets**

In this ablation experiment, we analyze the impact of the overlap between $J$ feature subsets on the model error and complexity. The configurations of main parameters of the TM and AMCRN model are as follows: $N=80$, $L=20$, $B=C=256$. We set three different overlaps between $J$ feature subsets, namely 0%, 25%, and 50%. The overlap of 0% means that there is no overlap between $J$ feature subsets and the input feature is split into four feature subsets ($J=80/20=4$). Similarly, the overlaps of 25% and 50% denote that there are 5-dimension overlap and 10-dimension overlap between $J$ feature subsets, respectively, and thus the input feature is divided into $J=5$ and $J=7$ feature subsets, respectively. Table III lists the scores of EER, PN, and MACs that are produced by the AMCRN model on the Vox1-O with three overlaps between $J$ feature subsets.

TABLE III
IMPACTS OF OVERLAP BETWEEN FEATURE SUBSETS ON THE MODEL ERROR AND MODEL COMPLEXITY WHEN EVALUATED ON THE VOX1-O

| Overlaps | $J$ | EER (%) | PN | MACs |
|---|---|---|---|---|
| 0% | 4 | 1.459 | 6.93 M | **0.362 G** |
| 25% | 5 | 1.442 | 6.93 M | 0.425 G |
| 50% | 7 | **1.391** | 6.93 M | 0.602 G |

Based on the results listed in Table III, we can obtain the following three observations.

First, as the overlap between feature subsets increases, the error decreases. For example, the EER score decreases from 1.459% to 1.391% when the overlap increases from 0% to 50%. The reason is probably that the more overlap between feature subsets, the richer correlation information contained in each feature subset, and the stronger the representational capability of speaker embedding. Hence, the model error can be lower.

Second, the memory requirement of the model is not related to the overlap between feature subsets. When the overlap increases from 0% to 50%, the PN value remains unchanged. The possible reason is described as follows. In our method, each feature subset is fed into the model independently. No matter how many feature subsets are, as long as the dimension of feature subsets and the model parameters remain unchanged, the PN value will keep the same.

Third, with the increase of the overlap between feature subsets, the computational complexity of the model increases. For example, the MACs value increases from 0.347 G to 0.602 G when the overlap increases from 0% to 50%. The reason is probably that the more overlap between feature subsets, the more feature subsets the model needs to process. Hence, the computational load of the model becomes heavier.

In conclusion, increasing the overlap between feature subsets will lead to the decrease of the model error, but it will result in the increase of the computational complexity of the model. Accordingly, in practical applications, we can set different overlaps between feature subsets as needed. For example, if there are only limits on the model size without restrictions on the computational complexity, the error can be reduced by reasonably increasing the overlap between feature subsets.

**Qualitative analysis for feature fusion**

In this ablation experiment, we aim to qualitatively analyze the impact of feature fusion on the representational capability of feature subsets and visualize this impact. The t-SNE [63] is one of the most popular methods for visualizing the similarities of different classes (samples). Although it cannot ensure that the distances between classes in low-dimensional space corresponds exactly to the distances in high-dimensional space, the distances between classes in high-dimensional space are well preserved in low-dimensional (e.g., two-dimensional) space [63]. Hence, we adopt the t-SNE to map the feature subsets $F_i$ and $F_i'$ (as shown in Fig. 2) into a two-dimensional space. $F_i$ and $F_i'$ are the input and output feature subsets of the feature fusion block of the TM, respectively. The *scikit-learn* (a Python library) is used to reduce the dimensionality of $F_i$ and $F_i'$, while the Python library of *matplotlib* is adopted to plot the distributions of both $F_i$ and $F_i'$ in the two-dimension space.

Without loss of generality, ten speakers are randomly chosen from the Vox1-O to show the distributions of their feature subsets $F_i$ and $F_i'$ in the two-dimensional space. We choose $F_i$ and $F_i'$ from the TM which is implanted into the AMCRN. We depict the distributions of $F_i$ and $F_i'$ ($1 \leq i \leq 4$) output from Block 4 of the AMCRN in two-dimensional space, as shown in Fig. 5. From the four subgraphs of Fig. 5, we can observe that the intervals between ten $F_i'$ are almost all larger than those between ten $F_i$. In other words, compared with the ten $F_i$ without feature fusion, the ten $F_i'$ with feature fusion are further apart in the two-dimensional space and the confusion between the ten $F_i'$ are expected to be less than that between the ten $F_i$.

Hence, the processing of the feature fusion block is helpful to enhance the representational capability of the transformed feature subsets. The stronger the representational capability of the feature subsets is, the smaller the error of the model will be.

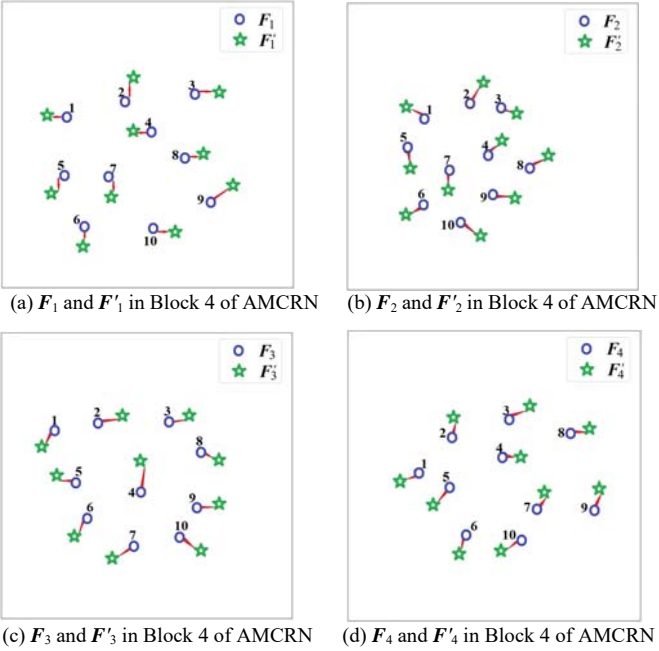

(a) $F_1$ and $F'_1$ in Block 4 of AMCRN  (b) $F_2$ and $F'_2$ in Block 4 of AMCRN
(c) $F_3$ and $F'_3$ in Block 4 of AMCRN  (d) $F_4$ and $F'_4$ in Block 4 of AMCRN

Fig. 5 Visualization of the $F_i$ and $F'_i$ ($1 \leq i \leq 4$) of ten speakers that are randomly selected from the Vox1-O. $F_i$ and $F'_i$ denote the input and output feature subsets of the feature fusion block of the TM, respectively. The digits from 1 to 10 represent ten different speakers.

### D. Comparison of Different Methods

In this section, we compare the proposed lightweight method to six state-of-the-art methods for lightweight SV, including the ECAPA-TDNNLite [50], EfficientTDNN [51], KD-based [52], Thin-ResNet34 [64], Fast-ResNet34 [65], and CSTCTS1dConv (Channel Split Time-Channel-Time Separable 1-dimensional Convolution) [66]. The ECAPA-TDNNLite based method [50] is a lightweight version of the ECAPA-TDNN based method, in which a large model, ECAPA-TDNN, is utilized for enrollment and a small model, ECAPA-TDNNLite, is used for verification. The EfficientTDNN based method [51] uses the NAS technique to design an efficient model to implement lightweight SV. The KD-based method [52] needs to train a large teacher model first, and then the KD technique is adopted to obtain a small student model based on the teacher model for realizing lightweight SV. The Thin-ResNet34 based method [64] and the Fast-ResNet34 based method [65] are lightweight versions of the ResNet34 based method. In the CSTCTS1dConv based method [66], a CSTCTS1dConv module was designed and then the KD technique was applied to improve the performance by learning better speaker embedding from the large model.

Main parameters of these baseline methods above are set according to the recommendations in corresponding references and then properly tuned on the training data. These baseline methods are implemented using the source codes released by the authors of these methods, or re-implemented by ourselves according to the descriptions in the corresponding references. Our proposed method for lightweight SV is implemented by implanting the proposed TM into three models and thus has three different versions, including the AMCRN-TM, the ResNet34-TM, and the ECAPA-TDNN-TM. Based on the introductions above, main technical characteristics between different methods for lightweight SV are briefly presented in Table IV.

TABLE IV
SUMMARY OF DIFFERENT METHODS FOR LIGHTWEIGHT SV

| Methods | Technical characteristics |
|---|---|
| ECAPA-TDNNLite | Lightening the ECAPA-TDNN |
| EfficientTDNN | Using the NAS to design model |
| KD-based | Using the KD to obtain teacher-student models |
| Thin-ResNet34 | Thin version of the ResNet34 |
| Fast-ResNet34 | Fast version of the ResNet34 |
| CSTCTS1dConv | Channel split time-channel-time separable 1D Conv |
| Ours | Implanting the plug-and-play TM into models |

Various methods are compared on three different evaluation sets (Vox1-O, Vox1-E, and Vox1-H) in model error (EER) and model complexity (including PN and MACs). Under the same experimental conditions, the scores of EER, PN and MACs that are achieved by various methods on three different evaluation sets are presented in Table V.

TABLE V
COMPARISON OF DIFFERENT METHODS IN TERMS OF EER, PN AND MACs

| Methods | EER (%) | | | PN | MACs |
|---|---|---|---|---|---|
| | Vox1-O | Vox1-E | Vox1-H | | |
| ECAPA-TDNNLite | 3.07 | 3.00 | 5.20 | 318.13K | 0.060G |
| EfficientTDNN | 2.20 | 2.37 | 3.79 | 0.90 M | 0.204G |
| KD-based | 2.638 | 2.729 | 4.117 | 970.01K | 0.081G |
| Thin-ResNet34 | 2.531 | 2.622 | 4.095 | 1.41 M | 5.403G |
| Fast-ResNet34 | 2.594 | 2.703 | 4.212 | 1.41 M | 2.712G |
| CSTCTS1dConv | 2.62 | 2.77 | 4.44 | 238.99K | 23.2M |
| AMCRN (1, 512) | 1.464 | 1.551 | 2.640 | 11.38 M | 0.840G |
| AMCRN-TM (2, 128) | 1.896 | 2.013 | 3.108 | 1.76 M | 0.093G |
| AMCRN-TM (4, 64) | 2.624 | 2.726 | 3.842 | 644.89K | 0.053G |
| AMCRN-TM (8, 32) | 2.981 | 3.108 | 4.219 | 360.79K | 0.043G |
| AMCRN-TM (16, 16) | 3.205 | 3.419 | 4.562 | 289.02K | 0.040G |
| ResNet34 (1, 512) | 1.674 | 1.752 | 3.440 | 23.91 M | 14.95G |
| ResNet34-TM (2, 128) | 2.015 | 2.107 | 2.812 | 9.32 M | 5.783G |
| ResNet34-TM (4, 64) | 2.799 | 2.915 | 4.639 | 3.82 M | 2.932G |
| ResNet34-TM (8, 32) | 3.144 | 3.216 | 5.014 | 1.97 M | 1.954G |
| ResNet34-TM (16, 16) | 3.446 | 3.573 | 5.317 | 1.20 M | 1.842G |
| ECAPA-TDNN (1, 512) | 1.010 | 1.241 | 2.320 | 6.20 M | 0.882G |
| ECAPA-TDNN-TM (2, 128) | 1.441 | 1.635 | 2.829 | 2.18 M | 0.090G |
| ECAPA-TDNN-TM (4, 64) | 1.922 | 2.138 | 3.218 | 0.89 M | 0.051G |
| ECAPA-TDNN-TM (8, 32) | 2.532 | 2.741 | 3.920 | 523.26K | 0.034G |
| ECAPA-TDNN-TM (16, 16) | 2.813 | 2.911 | 4.225 | 310.49K | 0.026G |

As shown in Table V, there are three versions of the proposed method and six baseline methods for performance comparison. For each version of the proposed method, we set five different numbers of feature subsets and channels, including (1, 512), (2, 128), (4, 64), (8, 32), and (16, 16). The left digit (i.e., 1, 2, 4, 8, 16) and the right digit (i.e., 512, 128, 64, 32, 16) in the brackets represent the number of feature subsets and the number of channels, respectively. The AMCRN (1, 512), ResNet34 (1, 512), and ECAPA-TDNN (1, 512) stand for the original method without the implantation of the TM. The AMCRN-TM (·, ·), the ResNet34-TM (·, ·), and the ECAPA-TDNN-TM (·, ·) represent the proposed methods for lightweight SV with different parameter configurations after the implantation of the TM. Based on the experimental results obtained by different methods in Table V, the following three conclusions can be drawn.

First, the complexity of three versions of the proposed method decreases significantly but their errors increase slightly, after the TM is implanted into the models of AMCRN, ResNet34 and ECAPA-TDNN. As for the AMCRN based method, the PN value decreases from 11.38 M (39.37 times of 289.02 K) to 289.02 K and the MACs value decreases from 0.840 G (21 times of 0.040 G) to 0.040 G. Meanwhile, the EER scores on the evaluation sets of Vox1-O, Vox1-E, and Vox1-H increase from 1.464% to 3.205% (2.19 times of 1.464%), from 1.551% to 3.419% (2.2 times of 1.551%), and from 2.640% to 4.562% (1.73 times of 2.640%), respectively. Similar results can be obtained for the ResNet34 based and ECAPA-TDNN based methods. That is, by setting proper parameters (e.g., numbers of channels and feature subsets), implanting the TM into the three models can remarkably reduce the model complexity with slight increase of the model error.

Second, three versions of the proposed method have different advantages in both error and complexity. The AMCRN-TM ( · , · ) based method has advantages over the methods based on the ECAPA-TDNN-TM ( · , · ) and the ResNet34-TM ( · , · ) in terms of PN. The ECAPA-TDNN-TM ( · , · ) based method outperforms the methods based on other two models in terms of EER and MACs. In addition, among the three versions of the proposed method, the ResNet34 based method has the heaviest computational load and the highest memory requirement. The reason is probably that compared with the AMCRN and the ECAPA-TDNN, the ResNet34 has the largest number of layers, and each layer (especially the convolutional layer) has a large amount of computation and parameters.

Third, the proposed methods outperform the six baseline methods on the whole in model error and model complexity. For example, the proposed methods of the AMCRN-TM (4, 64) and the ECAPA-TDNN-TM (16, 16) have advantages over the ECAPA-TDNNLite based method in terms of EER and MACs, and in terms of all metrics of EER, PN and MACs, respectively. The proposed methods of the ECAPA-TDNN-TM (4, 64) outperforms the baseline methods based on the EfficientTDNN, Thin-ResNet34 and Fast-ResNet3 in terms of all metrics of EER, PN and MACs. In addition, the proposed method of the AMCRN-TM (4, 64) exceeds the KD-based method in terms of all performance metrics of EER, PN and MACs. Although the CSTCTS1dConv based method achieves competitive results under similar model complexity, the proposed method of ECAPA-TDNN-TM (16, 16) beats it in EER on the Vox1-H.

In addition, the ResNet34-TM (16, 16) based method obtains a little higher EER scores than the baseline methods based on both the Thin-ResNet34 and the Fast-ResNet34. However, the model complexity (namely PN and MACs values) of the ResNet34-TM (16, 16) based method is much lower than that of the baseline methods based on the Thin-ResNet34 and the Fast-ResNet34. It should be noted that the goal of this work is not to achieve that the proposed method based on any models outperforms all existing lightweight SV methods in model error. Instead, our goal is that the proposed method using the plug-and-play TM can noticeably reduce the complexity of multiple kinds of models and meanwhile can keep the models' error basically unchanged.

In summary, the proposed TM can be easily implanted into several models (here taking three state-of-the-art models as examples) with different architectures without changing the structures of the original models. The proposed TM works in a plug-and-play way to noticeably reduce the complexity of multiple models rather than one specific model, and only leads to a slight increase in error. The remarkable difference between the proposed method and all existing methods is that the former is a general solution for lightening existing models for realizing lightweight SV, while the latter is a specific lightweight SV method. That is, we design a TM for reducing the complexity of many kinds of models, thereby achieving lightweight SV. However, all existing lightweight SV methods are designed for a specific model and thus lack generality. From the perspective of generality, the proposed method has advantage over existing lightweight SV methods.

### E. Robustness on Truncated Segments

In this section, we analyze the robustness of six baseline methods and three different representatives of our proposed method on the truncated testing segments. These methods have similar values of PN and MACs. We only give the scores of EER, because the value of PN has no relationship with the length of speech segments and the value of MACs definitely increases with the increase of the length of speech segments. For simplicity, we choose one representative from each of three versions of the proposed method, namely the AMCRN-TM (16, 16), the ResNet34-TM (16, 16), and the ECAPA-TDNN-TM (16, 16). The average length of each testing sample is about 8 seconds. Each truncated segment is obtained by randomly dividing each testing sample into segments with lengths of 2 seconds or 5 seconds. These truncated segments in the three different evaluation sets are adopted as the testing data to evaluate the robustness of various methods on the truncated testing segments with different lengths. Table VI presents the EER scores obtained by different methods on the truncated testing segments.

TABLE VI
EER SCORES (IN %) OBTAINED BY SIX BASELINE METHODS AND THREE REPRESENTATIVES OF OUR PROPOSED METHOD WHEN THEY ARE EVALUATED ON THE TRUNCATED TESTING SEGMENTS

| Methods | Length | Vox1-O | Vox1-E | Vox1-H |
|---|---|---|---|---|
| ECAPA-TDNNLite | 8 s | 3.07 | 3.00 | 5.20 |
|  | 5 s | 3.24 | 3.26 | 5.48 |
|  | 2 s | 5.03 | 5.12 | 7.29 |
| EfficientTDNN | 8 s | 2.20 | 2.37 | 3.79 |
|  | 5 s | 2.39 | 2.81 | 4.11 |
|  | 2 s | 4.22 | 4.60 | 6.02 |
| KD-based | 8 s | 2.638 | 2.729 | 4.117 |
|  | 5 s | 2.832 | 3.184 | 4.503 |
|  | 2 s | 4.641 | 4.873 | 6.311 |
| Thin-ResNet34 | 8 s | 2.531 | 2.622 | 4.095 |
|  | 5 s | 3.610 | 3.831 | 4.514 |
|  | 2 s | 5.491 | 5.772 | 6.963 |
| Fast-ResNet34 | 8 s | 2.594 | 2.703 | 4.212 |
|  | 5 s | 3.675 | 3.922 | 4.635 |
|  | 2 s | 5.571 | 5.864 | 7.087 |
| CSTCTS1dConv | 8 s | 2.62 | 2.77 | 4.44 |
|  | 5 s | 2.91 | 3.19 | 4.72 |
|  | 2 s | 4.63 | 4.88 | 6.51 |
| AMCRN-TM (16, 16) | 8 s | 3.205 | 3.419 | 4.562 |
|  | 5 s | 3.341 | 3.533 | 4.738 |
|  | 2 s | 5.013 | 5.104 | 6.714 |
| ResNet34-TM (16, 16) | 8 s | 3.446 | 3.573 | 5.317 |
|  | 5 s | 4.619 | 4.771 | 5.723 |
|  | 2 s | 6.502 | 6.763 | 8.143 |
| ECAPA-TDNN-TM (16, 16) | 8 s | 2.813 | 2.911 | 4.225 |
|  | 5 s | 2.957 | 3.224 | 4.502 |
|  | 2 s | 4.516 | 4.787 | 6.269 |

Based on the experimental results produced by six baseline methods and the three representatives of our proposed method in Table VI, we can draw the following two conclusions.

First, the EER scores achieved by all methods on testing segments steadily increase with the decrease of the segment length. Furthermore, the increase of EER scores obtained by the three proposed methods is smaller than that obtained by most baseline methods. For instance, when the lengths of speech segments in the Vox1-H decrease from 8 seconds to 2 seconds, the absolute increment of the EER score achieved by the proposed method of ECAPA-TDNN-TM (16, 16) is 2.044% (6.269% - 4.225%). This value (2.044%) is smaller than the counterparts obtained by all baseline methods.

Second, as the length of the testing segments decreases, the proposed methods gradually achieve lower EER scores than most baseline methods. For example, the proposed methods of ECAPA-TDNN-TM (16, 16) and AMCRN-TM (16, 16) obtains the EER scores of 2.813% and 3.205%, respectively, when they are evaluated on the testing segments of 8 seconds in Vox1-O. These two EER scores of 2.813% and 3.205% are higher than the EER scores of 2.62% (obtained by the method of CSTCTS1dConv) and 2.531% (obtained by the method of Thin-ResNet34), respectively. However, when these methods are assessed on the testing segments of 2 seconds in Vox1-O, the proposed methods of ECAPA-TDNN-TM (16, 16) and AMCRN-TM (16, 16) obtains lower EER scores than the baseline methods of CSTCTS1dConv and Thin-ResNet34, respectively. Similar results can be obtained when the proposed methods and baseline methods are evaluated on the testing segments in other evaluation sets.

In conclusion, compared with the baseline methods, the EER scores of the proposed methods are less affected by the length of testing segments. That is, the proposed methods are more robust than baseline methods on truncated testing segments. Accordingly, the proposed methods can generalize well across truncated testing segments with different lengths instead of overfitting on testing segments with fixed length.

## VI. Conclusions

In this study, we focused on solving the problem of lightweight SV with both lower complexity and lower error. To realize this goal, we designed a TM to conduct feature partition and fusion on the input feature. Afterwards, we proposed a method for lightweight SV by implanting the proposed TM into three state-of-the-art models with different architectures. Based on the description of both the designed TM and the proposed method for lightweight SV, and the experimental discussions, the following two conclusions can be drawn.

First, the proposed TM worked well for reducing the model complexity while obtaining equivalent or even lower error for SV. Moreover, it can be implanted into the models with different architectures in a plug-and-play way without the need to change the structures of the original models. On the contrary, the baseline methods for model lightweight need to change the models by redesigning the entire architectures or replacing some modules (blocks) of the original models. In addition, the proposed TM can be used to reduce the complexity of multiple types of models, whereas each one of the baseline methods is generally effective for one specific model.

Second, the proposed method for lightweight SV exceeded the state-of-the-art methods on the whole in terms of EER, PN and MACs, when evaluated on three different evaluation sets. In addition, the proposed method generalized well on truncated testing segments with various lengths.

Although our proposed method has achieved encouraging results for lightweight SV, there are still some aspects to be improved in it. First, the application scope of the proposed TM needs to be extended. Due to the fact that 1D-convolutional blocks are adopted in most state-of-the-art models (e.g., ECAPA-TDNN) for SV, the convolutional operations in the proposed TM are designed to be 1D (see Fig. 2). Hence, the proposed TM is effective for lightening the 1D-convolutional models only. In future work, we will design a TM that includes 2D-convolutional blocks and use the proposed TM to lighten the 2D-convolutional models. We will also consider designing a general TM that can effectively lighten the models of any structures, such as the models including 1D-convolutional and 2D-convolutional blocks. Second, we discussed the lightweight of three state-of-the-art models only. We basically placed the proposed TM in front of the blocks of the original models, but not in each layer of these blocks. In future work, we will investigate the lightweight variants of other types of models, and explore the implantation of the proposed TM in front of all layers of the original models. Third, we did not investigate the implementation of the proposed method on intelligent speech terminals with limited resources. In next work, we will consider the implementation of the proposed method on the portable speech terminals for the forensic scenarios. To achieve this goal, we will further reduce the model complexity and keep the model error as lower as possible by taking effective measures. For example, we will optimize the structure of the proposed TM, and integrate other techniques (e.g., model quantization) into our proposed method.